\documentstyle[aps,preprint,eqsecnum,tighten,floats,epsf,rotate]{revtex}

\newbox\rotbox

\begin{document}
\draft
\setcounter{page}{0}
\def\footnoterule{\kern-3pt \hrule width\hsize \kern3pt}
\title{WEAK HYPERON PRODUCTION IN $ep$ SCATTERING\thanks
{This work is supported in part by funds provided by the U.S.
Department of Energy (D.O.E.) under cooperative 
research agreement \#DF-FC02-94ER40818.}}

\author{Xuemin Jin
\footnote{Email address: {\tt jin@ctp02.mit.edu}}
and R. L. Jaffe
\footnote{Email address: {\tt jaffe@mitlns.mit.edu}}
}

\address{Center for Theoretical Physics \\
Laboratory for Nuclear Science \\ 
and Department of Physics \\
Massachusetts Institute of Technology \\
Cambridge, Massachusetts 02139 \\}

\date{MIT-CTP-2594,~ hep-th/9612316. {~~~~~} December 1996}
\maketitle

\thispagestyle{empty}

\begin{abstract}

We study the kinematics and cross section of the scattering process 
$e  p \rightarrow e  \Sigma^+$. The cross section is expressed in terms 
of complex form factors characterizing the hadron
vertices. We estimate the cross section for small momentum transfer using 
known experimental information. To first order in the momentum transfer,
we obtain a model independent result for the photon-exchange part
of the cross section, which is completely determined by the decay width 
$\Gamma(\Sigma^+ \rightarrow p\gamma)$. For the kinematics of the parity 
violation experiment at MAMI, this first order result gives rise to a ratio 
of $(d\sigma/d\Omega)_{ep\rightarrow e\Sigma^+}$ $/(d\sigma/d\Omega)_{ep\rightarrow ep} 
\simeq 4.0\times 10^{-15}$. The $Z^0$-exchange and interference parts give 
much smaller contributions due to the suppression of the flavor changing
weak neutral current in the standard model. Feasibility of the
experimental measurement is briefly discussed.

\end{abstract}

\vspace*{\fill}
\begin{center}
Submitted to: {\it Physical Review D}
\end{center}
\narrowtext
\section{Introduction}

Currently, experimentalists at MAMI in Mainz are considering the
possibility of measuring the scattering processes 
$e p\rightarrow e \Sigma^+$ or $e d \rightarrow e p \Lambda$~\cite{maas}. 
To our knowledge, there have been no theoretical investigations of 
such processes. It is thus important to explore the physics 
and the feasibility of the experimental measurement. 
Here we discuss the kinematics and cross section for $e p \rightarrow e \Sigma^+$. 
With minor alterations our analysis applies as well to $e n\rightarrow e \Lambda$.

In lowest order the scattering process $e p \rightarrow e \Sigma^+$ proceeds 
via the exchange of one photon or one $Z^0$ boson (see Fig.~\ref{fig-1}). 
Thus, the cross section consists of a pure $\gamma$, pure $Z^0$ and interference 
parts. The last two depend on the physics at the 
$pZ^0\Sigma^+$ vertex, a classic flavor changing weak
neutral current, which is severely suppressed in the standard model.
On the other hand, the photon-exchange part,
$p \gamma \Sigma^+$, includes all gauge interactions of standard model: 
strong, weak, and electromagnetic. The same vertex appears in
the weak radiative decay $\Sigma^+\rightarrow p \gamma$, which has been 
well measured experimentally~\cite{pdt}. 
Despite substantial theoretical effort~\cite{comment}, 
hyperon weak radiative decays remain
poorly understood. Measurement of $e p \rightarrow e \Sigma^+$ might
provide more information about the vertex $p \gamma \Sigma^+$
and hence constrain theoretical models.

We express the cross section in terms of various form factors, 
which reflect the structure of the hadron vertices. We then estimate the 
cross section at small four-momentum transfer using known experimental 
information on the weak radiative decay, $\Sigma^+ \rightarrow p \gamma$, and
the flavor changing weak neutral current. To first order in the momentum 
transfer, we obtain a model independent result for the photon-exchange part, 
which is completely determined by the weak radiative decay width 
$\Gamma(\Sigma^+\rightarrow p\gamma)$.  For the kinematics of the parity 
violation experiment at MAMI ($\theta \simeq 35^0$ and 
$q^2\simeq -0.237$ GeV$^2$ for the elastic scattering $e p \rightarrow e p$, 
with $\theta$ the scattering angle and $q^2$ the squared four-momentum 
transfer)~\cite{maas96}, this first order result leads to a suppression
factor of $4\times 10^{-15}$ relative to the $ep$ elastic scattering. 
The physical reasons for the suppression 
are the factor of $G_F^2$ from the weak hamiltonian {\it and\/} a factor
of $q^2$ from electromagnetic gauge invariance that suppresses the cross
section at small momentum transfer.   Using the experimental result for the
branching ratio of 
$K^0_L\rightarrow \mu^+\mu^-$, we estimate that the $Z^0$-exchange and 
interference parts give negligible contributions. We shall discuss briefly 
the feasibility of experimental measurements at available facilities.

This paper is organized as follows: In Sec.~\ref{II}, we discuss the
kinematics and derive the cross section. Sec.~\ref{III} gives an estimate
of the cross section. Sec.~\ref{IV} is devoted to summary and conclusion.

\section{Kinematics and cross section}
\label{II}

\begin{figure}[b]
\begin{center}
\epsfysize=7.0truecm
\leavevmode
\epsfbox{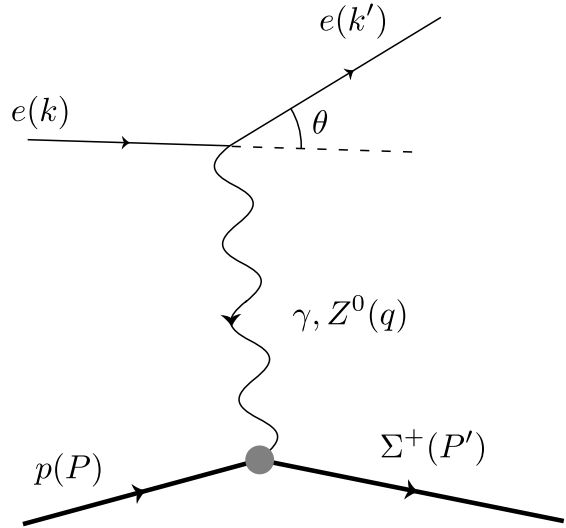}
\end{center}
\caption{Kinematics of the scattering  $e p\rightarrow e \Sigma^+$ in the
proton rest frame.}
\label{fig-1}
\end{figure}

The kinematics for the process $e p\rightarrow e \Sigma^+$ is illustrated 
in Fig.~\ref{fig-1}. The four momentum of the initial and final states are denoted 
by $k = \{E,{\bf k}\}$ for the initial electron, $P = \{E_p,{\bf p}\}$ for 
the target (proton) ($P = \{M_p,{\bf 0}\}$ in the target rest frame), 
$k^\prime = \{E^\prime, {\bf k^\prime}\}$ for the outgoing electron, and 
$P^\prime = \{E_p^\prime, {\bf p^\prime}\}$ for the outgoing hyperon ($\Sigma^+$). 
One can then define the two invariants
\begin{eqnarray}
q^2& \equiv & (k - k^\prime)^2 = (P^\prime - P)^2 
= -4 E E^\prime \sin^2(\theta/2) = - Q^2 < 0\ ,
\label{invar-q}
\\*[7.2pt]
\nu& \equiv & P\cdot q = M_p (E - E^\prime)\ ,
\label{invar-pq}
\end{eqnarray}
where the electron mass has been neglected (and will be henceforth).
The scattering angle $\theta$ has been indicated in Fig.~1. Unless otherwise
noted, the target rest frame will be adopted. 
%
%

Since the final $\Sigma^+$ state is on-shell, one finds
\begin{eqnarray}
& & 2 \nu + q^2 = M_{\Sigma}^2 - M_p^2\ ,
\label{nu-q}
\\*[7.2pt]
& & E^\prime = {M_p E - {1\over 2} (M_\Sigma^2 - M_p^2)
\over M_p + 2 E \sin^2(\theta/2)}\ .
\label{eprime}
\end{eqnarray}
Note that there are only two independent variables, $E$ and $\theta$, and
all the other quantities can be expressed in terms of them.
The kinematic domain is thus given by the two conditions $q^2 < 0$ $(Q^2 > 0)$ 
and $2\nu = M_\Sigma^2 - M_p^2 - q^2$. Since $E^\prime > 0$, one has
\begin{equation}
E > {1\over 2 M_p} \left(M_\Sigma^2 - M_p^2\right)\ ,
\end{equation}
which implies a minimum initial electron energy of 
$E_{\rm min} \simeq 285$ MeV.

In the discussions to follow, we shall consider the scattering process
with unpolarized beam and target and with final spins unobserved. 
The extension to other situations is straightforward. The differential 
cross section in this case can be written as:
\begin{equation}
d \sigma = {e^4\over 4 E Q^4}
\int {d^3 k^\prime\over (2\pi)^3 E^\prime}
{d^3 p^\prime \over (2\pi)^3}{M_\Sigma\over E_p^\prime} 
(2\pi)^4 \delta^4 (P^\prime + k^\prime - P - k) 
\sum_{i=\gamma,\gamma Z,Z} \eta^i l^{\mu\nu}_i W^i_{\mu\nu}\ ,
\label{dsigma}
\end{equation}
where 
\begin{equation}
\eta^\gamma = 1\ ,
\hspace*{1cm}
\eta^{\gamma Z} = {1\over \sin^2 2\theta_{\rm w}}\,
{Q^2\over Q^2+M_Z^2}\ ,
\hspace*{1cm}
\eta^Z = \left( \eta^{\gamma Z}\right)^2\, ,
\label{coupling-f}
\end{equation}
with $\theta_{\rm w}$ the weak angle and $M_Z$ the $Z^0$ mass. 
Here we use the normalization conventions of Itzykson and Zuber~\cite{book}.
The leptonic tensor $l^{\mu\nu}_i$ is simply given by
\begin{eqnarray}
 l^{\mu\nu}_\gamma &=& k^\mu k^{\prime\nu} + k^\nu k^{\prime\mu} - 
k\cdot k^\prime g^{\mu\nu}\ ,
\\*[7.2pt]
 l^{\mu\nu}_{\gamma Z} &=& 2 g^e_{\rm v} 
\left (k^\mu k^{\prime\nu} + k^\nu k^{\prime\mu} - 
k\cdot k^\prime g^{\mu\nu} \right)
-2 g^e_{\rm a} i \epsilon^{\mu\nu\rho\sigma}k^\prime_\rho k_\sigma\ ,
\\*[7.2pt]
 l^{\mu\nu}_Z &=& \left[ (g^e_{\rm v})^2 +(g^e_{\rm a})^2\right]
\left (k^\mu k^{\prime\nu} + k^\nu k^{\prime\mu} - 
k\cdot k^\prime g^{\mu\nu} \right)
-2 g^e_{\rm v} g^e_{\rm a} i \epsilon^{\mu\nu\rho\sigma}k^\prime_\rho k_\sigma\ ,
\label{lmn}
\end{eqnarray}
where 
\begin{equation}
g^e_{\rm v} = -{1\over 2} + 2 \sin^2\theta_{\rm w}\ , \hspace*{1cm}
g^e_{\rm a} = -{1\over 2}\ .
\label{lepton-coupling}
\end{equation}
On the other hand, the hadronic tensor $W_{\mu\nu}^i$ describes the complicated 
structure of the vertices $p \gamma \Sigma^+$ and $p Z^0\Sigma^+$, 
\begin{eqnarray}
W_{\mu\nu}^\gamma &=&\sum_{\rm spins}
\langle \Sigma^+| J_\mu^\gamma |p\rangle 
\langle \Sigma^+| J_\nu^\gamma |p\rangle^*\ ,
\\*[7.2pt]
W_{\mu\nu}^{\gamma Z} &=& \sum_{\rm spins}\left[
\langle \Sigma^+| J_\mu^\gamma |p\rangle
\langle \Sigma^+| J_\nu^Z |p\rangle^*
+\langle \Sigma^+| J_\mu^Z |p\rangle
\langle \Sigma^+| J_\nu^\gamma |p\rangle^*\right]\ ,
\\*[7.2pt]
W_{\mu\nu}^Z &=& \sum_{\rm spins}
\langle \Sigma^+| J_\mu^Z |p\rangle 
\langle \Sigma^+| J_\nu^Z |p\rangle^*\ ,
\label{w-def}
\end{eqnarray}
where $J_\mu^\gamma$ and $J_\mu^Z$ denote the electromagnetic and weak 
neutral currents, respectively. The matrix element
$\langle \Sigma^+| J_\mu^j |p\rangle$ $\{j = \gamma, Z\}$, 
upon the use of Lorentz covariance, takes the following form:
\begin{eqnarray}
\langle \Sigma^+ |J_\mu^j |p\rangle &\equiv&
\overline{U}(P^\prime)
\Bigl\{\gamma_\mu F_1^j(q^2)
+ q_\mu B^j(q^2)+{i \sigma_{\mu\nu} q^\nu
\over M_p + M_\Sigma} F_2^j (q^2)
\nonumber
\\*[7.2pt]
& & + \gamma_5\gamma_\mu F_3^j(q^2) + \gamma_5 q_\mu B_5^j(q^2)
+ {i\gamma_5\sigma_{\mu\nu}q^\nu\over M_p - M_\Sigma} F_4^j(q^2)
\Bigr\} U(P)\ ,
\label{vertex-gen}
\end{eqnarray}
where $U(P^\prime)$ and $U(P)$ are the Dirac spinors of $\Sigma^+$ and proton,
respectively. The form factors are in general complex.
While $F_1^j, F_2^j$, and $B^j$
are parity-conserving, $F_3^j, F_4^j$, and $B^j_5$ are parity-violating.  

From electromagnetic current conservation, one obtains the 
following relations:
\begin{equation}
(M_p - M_\Sigma)\, F_1^\gamma = q^2 B^\gamma\ , 
\hspace*{1.5cm} (M_p + M_\Sigma)\,  F_3^\gamma = q^2 B_5^\gamma\ .
\label{curr-cons}
\end{equation}
Since there is no zero-mass particle involved except the photon and 
the photon propagator has been included explicitly, Eq.~(\ref{curr-cons})
implies that
\begin{equation}
F_1^\gamma(0) = 0, \hspace*{2cm}F_3^\gamma(0) = 0\ .
\label{cons-fs}
\end{equation}
Note that the condition $F_1^\gamma(0) = 0$ is different from the usual 
$F_1^\gamma(0) = 1$ (= electric charge of the proton) seen in the proton 
electromagnetic form factors. This difference arises from gauge invariance. 
While the proton couples minimally to the photon field $A_\mu$,  
the $p\gamma \Sigma^+$ vertex must be proportional to
$F_{\mu\nu} = \partial_\mu A_\nu - \partial_\nu A_\mu$.
Thus, there are only four independent form factors at the vertex 
$p \gamma \Sigma^+$, which we choose as $F_1^\gamma, F_2^\gamma, F_3^\gamma$, 
and $F_4^\gamma$. Since $J^Z_\mu$ is not conserved,
the six form factors describing the vertex $p Z^0\Sigma^+$ are
in general independent.

Lorentz covariance and electromagnetic current conservation 
allows one to separate $W_{\mu\nu}^\gamma$ into three distinct structures:
\begin{eqnarray}
W_{\mu\nu}^\gamma
& \equiv & \left( -g_{\mu\nu} 
+ {q_\mu q_\nu\over q^2}\right) {M_p\over M_\Sigma}  W_1^\gamma(q^2)
+\left[\left( P_\mu - {P\cdot q\over q^2} q_\mu\right) 
\left(P_\nu - {P\cdot q\over q^2}q_\nu\right)\right]
{W_2^\gamma(q^2)\over M_p M_\Sigma}
\nonumber
\\*[7.2pt]
& & -i \epsilon_{\mu\nu\rho\sigma} P^\rho q^\sigma 
{W_3^\gamma(q^2)\over 2 M_p M_\Sigma}\ .
\label{w-g-decop}
\end{eqnarray}
Here the introduction of the factors $M_p/ M_\Sigma$, $1/M_p M_\Sigma$, and
$1/2 M_p M_\Sigma$ makes the $W^\gamma$'s dimensionless. On the other hand,
$W_{\mu\nu}^i$ for $\{i=\gamma Z, Z\}$ can be decomposed into six
structures:
\begin{eqnarray}
W_{\mu\nu}^i
& \equiv & -g_{\mu\nu} {M_p\over M_\Sigma} W^i_1(q^2) 
+P_\mu P_\nu {W_2^i(q^2)\over M_p M_\Sigma}
  -i \epsilon_{\mu\nu\rho\sigma} P^\rho q^\sigma 
{W_3^i(q^2)\over 2 M_p M_\Sigma}
+{\cal O}(q_\mu \, \mbox{or} \, q_\nu)\ .
\label{w-z-decop}
\end{eqnarray}
%
Here terms proportional to $q_\mu$ or $q_\nu$ vanish upon contraction with
$l_i^{\mu\nu}$. Carrying out the integral in Eq.~(\ref{dsigma}), we obtain 
the following expressions for the three parts of the cross section:
\begin{eqnarray}
\left({d\sigma \over d\Omega}\right)_\gamma
&=&
\left({d\sigma \over d\Omega}\right)_0 \eta^\gamma
\left[ W_2^\gamma (q^2) + 2 \tan^2\left(\theta/ 2\right) W_1^\gamma (q^2)
\right]\ ,
\label{sig-gamma}
\\*[7.2pt]
\left({d\sigma \over d\Omega}\right)_{\gamma Z}
&=&
\left({d\sigma \over d\Omega}\right)_0 \eta^{\gamma Z}\, 
\Bigl\{ 2g^e_{\rm v}\left[ W_2^{\gamma Z} (q^2) 
+ 2 \tan^2\left(\theta/ 2\right) W_1^{\gamma Z} (q^2)
\right] 
\nonumber
\\*[7.2pt]
& & \hspace*{3.5cm}
-2 g^e_{\rm a} {E+E^\prime\over M_p}\tan^2 \left(\theta/ 2\right) 
W_3^{\gamma Z} (q^2)\Bigr\}\ ,
\label{sig-gammaz}
\\*[7.2pt]
\left({d\sigma \over d\Omega}\right)_Z
&=&
\left({d\sigma \over d\Omega}\right)_0 \eta^Z\, 
\Bigl\{ \left[(g^e_{\rm v})^2+(g^e_{\rm a})^2\right] \left[ W_2^Z (q^2) 
+ 2 \tan^2\left(\theta/ 2\right) W_1^Z (q^2)
\right] 
\nonumber
\\*[7.2pt]
& & \hspace*{5cm}
-2 g^e_{\rm v} g^e_{\rm a} 
{E+E^\prime\over M_p}\tan^2 \left(\theta/2\right) 
W_3^Z (q^2)\Bigr\}\ ,
\label{sig-z}
\end{eqnarray}
where
\begin{equation}
\left({d\sigma \over d\Omega}\right)_0
= {\alpha_E^2 
\over 4 E^2 \sin^4(\theta/2)}
\left\{{\cos^2\left(\theta/ 2\right)\over 
2 [1+2 (E/M_p)\sin^2(\theta/2)]}\right\}\ ,
\label{sig-0}
\end{equation}
with $\alpha_E$ the electromagnetic fine structure constant.
Therefore, the cross section is determined by various $W^j_k$,
which, in terms of the form factors, can be expressed as:
\begin{eqnarray}
W_1^j & = & {1\over 2M_p^2}\left\{ |F_1^j+F_2^j|^2
\left[Q^2 + \left(M_p - M_\Sigma\right)^2\right]
+|F_3^j+F_4^j|^2\left[Q^2 + \left(M_p + M_\Sigma\right)^2\right]
\right\}\ ,
\label{w1-gamma}
\\*[7.2pt]
W_2^j & = & 2 \left\{ |F_1^j|^2 +|F_3^j|^2 +Q^2 
\left[{|F_2^j|^2\over (M_p + M_\Sigma)^2}
+ {|F_4^j|^2 \over (M_p - M_\Sigma)^2}\right]
\right\}\ ,
\label{w2-gamma}
\\*[14.4pt]
W_3^j & = & -4\, {\rm Re}\left[(F_1^j+F_2^j) (F_3^j+F_4^j)^*\right]\ ,
\label{w3-gamma}
\end{eqnarray}
for $\{j = \gamma, Z\}$, where ${\rm Re}$ denotes the real part. 
The expressions for $W_k^{\gamma Z}$, which contain the product of 
$F_k^\gamma$ and $F_k^Z$, have not been listed.
\section{Estimate of the cross section}
\label{III}

The hadron vertices $p \gamma \Sigma^+$ and $pZ^0\Sigma^+$ are very complicated, 
containing the interplay among the strong, weak, and (for
$p\gamma\Sigma^+$)  electromagnetic  interactions.  Obviously, it is difficult to
calculate the form factors and cross  section directly from the standard model.
At this stage, the best one could do is  to calculate the form factors in
models. For the experiment conditions of interest we find it
possible to calculate the cross section without invoking  any explicit effective
model, by making use of experimental information on the weak  radiative decay
$\Sigma^+ \rightarrow p \gamma$ and the flavor changing weak  neutral current.

Let us first consider the photon-exchange part. It is easy to show that in  
our notation, the weak radiative decay width of the $\Sigma^+$ can 
be written as
\begin{equation}
\Gamma (\Sigma^+ \rightarrow p \gamma) 
= {e^2\over \pi} \left({M_\Sigma^2 - M_p^2\over 2 M_\Sigma}\right)^3
\left[ {|F_2^\gamma(0)|^2\over (M_p + M_\Sigma)^2} +
{|F_4^\gamma(0)|^2\over (M_p - M_\Sigma)^2 }\right]\ ,
\label{w-decay}
\end{equation}
and the asymmetry parameter as
\begin{equation}
\alpha_\gamma = 2 \left( {M_p + M_\Sigma \over M_p - M_\Sigma} \right)
\left\{ {{\rm Re} \left[F_2^\gamma(0) F_4^{^\gamma *} (0)\right]\over 
|F_2^\gamma(0)|^2 + 
\left[(M_p + M_\Sigma)/(M_p - M_\Sigma)\right]^2 |F_4^\gamma(0)|^2}\right\}\ .
\label{asy-par}
\end{equation}
Experimentally, the branching ratio of $\Sigma^+ \rightarrow p \gamma$
is $(1.25\pm 0.07) \times 10^{-3}$, and 
$\alpha_\gamma = -0.76\pm 0.08$~\cite{pdt}. 
Note that $\tan^2(\theta/2)\sim q^2$ [Eq.~(\ref{invar-q})], 
and $F_1^\gamma(q^2)\sim q^2$, $F_3^\gamma(q^2)\sim q^2$ [Eq.~(\ref{cons-fs})]
(at most) for small $q^2$. To first order in $q^2$, we have
\begin{eqnarray}
W_2^\gamma(q^2) &+& 2\tan^2\left({\theta\over 2}\right) W_1^\gamma(q^2)
= -2 q^2 \left[ {|F_2^\gamma(0)|^2\over (M_p + M_\Sigma)^2}
+ {|F_4^\gamma(0)|^2\over (M_p - M_\Sigma)^2} \right]
\nonumber
\\*[7.2pt]
& +& 2\tan^2\left({\theta\over 2}\right)
\left[{(M_p^2 - M_\Sigma^2)^2\over 2M_p^2}\right]
\left[{|F_2^\gamma(0)|^2\over (M_p + M_\Sigma)^2}
+ {|F_4^\gamma(0)|^2\over (M_p - M_\Sigma)^2}\right]  
+ {\cal O}[(q^2)^2]
\nonumber
\\*[7.2pt]
& =&  -(2 q^2)\times 3.11\times 10^{-14} ({\rm GeV})^{-2}
+ 2\tan^2\left({\theta\over 2}\right)\times 5.05\times 10^{-15}
+ {\cal O}[(q^2)^2]\ .
\label{cross-expan}
\end{eqnarray}
Here in the last step, we have used the experimental values 
for $\Gamma(\Sigma^+ \rightarrow p\gamma)$ and the lifetime of 
$\Sigma^+$. Therefore, the above first order result is model
independent. To be concrete, we consider 
the kinematics of the parity violation experiment at MAMI~\cite{maas96}, 
$\theta = 35^0$ and $q^2 \simeq -0.237$ GeV$^2$ for the elastic scattering, 
which implies $q^2 \simeq -0.16$ GeV$^2$ for the process under consideration.
Using the above estimate, we arrive at the following result:
\begin{equation}
{\left({\textstyle d\sigma\over \textstyle  d\Omega}\right)_\gamma 
\left(e p\rightarrow e \Sigma^+\right)
\over {\textstyle d\sigma\over \textstyle  d\Omega}
\left(e p \rightarrow e p\right)}
\simeq 4\times 10^{-15}\ .
\label{result}
\end{equation}
This indicates a severe suppression relative to elastic 
scattering. This suppression arises from the weak interaction, 
gauge invariance, and kinematics. 

We observe from Eqs.~(\ref{w1-gamma}) and (\ref{w2-gamma}) that $W_1^\gamma$
and $W_2^\gamma$ may increase as $Q^2$ gets larger, implying, perhaps, a
larger value for the ratio of $(d\sigma/d\Omega)_{ep\rightarrow
e\Sigma^+}/(d\sigma/d\Omega)_{ep\rightarrow ep}$. On the other 
hand, as $Q^2$ goes to infinity, one expects all the form
factors go to zero. Thus, there may be a chance that the ratio
has a maximum at a non-zero momentum transfer, which, however, is unlikely
to alter the result of Eq.~(\ref{result}) qualitatively.

The $Z^0$-exchange part and the interference part involve the flavor changing 
weak neutral current, which is suppressed (at the tree level) in the standard 
model. At the quark level, the relevant vertex is $s Z^0 d$. For the
purposes at hand, we can parameterize  this vertex in terms of a vector
coupling $g^Z_{\rm V}$ and an axial vector coupling
$g^Z_{\rm A}$. Note that the same vertex is also responsible for the decay 
$K^0_L\rightarrow \mu^+\mu^-$. Although the typical momentum scales in the
two processes ($ep\rightarrow e\Sigma^+$ and $K_L\rightarrow \mu^+\mu^-$)
differ by multiples of some typical hadronic scale, we will not make a
significant error by treating $g^Z_{\rm V}$ and $g^Z_{\rm A}$ as
constants. Since for electron and muon, $g^{e,\mu}_{\rm v}\sim 0$, 
only $g^Z_{\rm A}$ enters.

Thus we can make use of the experimental information on the 
decay $K^0_L\rightarrow \mu^+\mu^-$ to estimate $g^Z_{\rm A}$. 
Since the decay width of $K^0_L\rightarrow \mu^+\mu^-$ agrees with the 
standard model estimate (via $K^0_L\rightarrow \gamma\gamma \rightarrow \mu^+\mu^-$),
we can safely assume $g^Z_{\rm A}$ contribution to $K^0_L$ decay to be below the limit
of the experimental errors on the $K^0_L\rightarrow \mu^+\mu^-$ branching 
ratio. This gives:
\begin{equation}
|g^Z_{\rm A}| \leq {1\over 2}\sin\theta_{\rm c} \cos\theta_{\rm w}
\left[{\Gamma (K^0_L\rightarrow \mu^+\mu^-) \over
\Gamma (K^+\rightarrow \mu^+\nu_\mu)}\right]^{1/2}
\simeq 2\times 10^{-6}\ ,
\label{fnc3-est}
\end{equation}
where $\theta_{\rm c}$ is the Cabibbo angle. Here we have neglected
phase space difference, which is expected to be small. 

With the neglect of QCD binding effects, we can apply the above estimate
directly to the vertex $p Z^0\Sigma^+$, where only the axial vector coupling
contributes at $Q^2\sim 0$. We then obtain
\begin{equation}
{\left({\textstyle d\sigma\over \textstyle  d\Omega}\right)_Z
\left(e p\rightarrow e \Sigma^+\right)
\over {\textstyle d\sigma\over \textstyle  d\Omega}
\left(e p \rightarrow e p\right)}
\leq 7\times 10^{-22}\ .
\label{est-z}
\end{equation}

Therefore, the cross section for the process $e p \rightarrow e\Sigma^+$ is 
dominated by the photon exchange part. The $Z^0$-exchange and the interference 
give negligible contributions because of the suppression of the flavor changing 
weak neutral current.

\section{Summary and conclusion}
\label{IV}

In this paper, we have studied the kinematics and derived the cross
section of the scattering process $e p \rightarrow e \Sigma^+$. The 
cross section can be expressed in terms of various form factors
which are introduced to parameterize the physics at the vertices
$p \gamma \Sigma^+$ and $pZ^0\Sigma^+$. We have estimated the cross section 
at small momentum transfer invoking known experimental information. 
In particular, we obtain a model independent first order (in momentum transfer) 
result for the photon-exchange part of the cross section, which is completely 
determined by the weak radiative decay width $\Gamma(\Sigma^+\rightarrow p\gamma)$. 
With this result and the kinematics of the parity violation experiment at MAMI, 
we found a factor of $4\times 10^{-15}$ suppression relative to the 
elastic process $e p\rightarrow e p$. This suppression largely results from 
the weak interaction involved and the gauge invariance at the vertex $p\gamma\Sigma^+$. 
The $Z^0$-exchange part and the interference part both depend on the flavor 
changing weak neutral current. Using the experimental result for the branching
ratio of $K^0_L\rightarrow \mu^+\mu^-$, we estimate that these two parts are much 
smaller than the photon-exchange contribution.

In principle, the weak hyperon production process 
$e p\rightarrow e \Sigma^+$ discussed here can be measured 
experimentally in facilities with low energy electron beams.
For low momentum transfer, MAMI, MIT-Bates, and TJNAF
are the best candidates. However, our estimate shows 
that the cross section is severely suppressed relative to
the $ep$ elastic scattering. To be specific, 
consider the parity violation experiment proposed at MAMI.
There, about $10^{14}$ elastic events can be accumulated
(in 700 hours)~\cite{maas96}, making a meaningful
measurement unlikely at this time.


\acknowledgements

We are indebted to Frank Maas for suggesting the scattering 
process addressed here to us and providing useful information
on the parity violation experiment at MAMI. We would like to thank 
Lisa Randall for useful conversations.


\end{document}